\pdfoutput=1
\documentclass[aps,prd,reprint,superscriptaddress,longbibliography,nofootinbib]{revtex4-1}
\usepackage{graphicx}
\usepackage{amsmath}
\usepackage{amssymb}
\usepackage{amsfonts}
\usepackage{dcolumn}
\usepackage{dsfont}
\usepackage{latexsym}
\usepackage{rotating}
\usepackage{color}
\usepackage{latexsym}
\usepackage{bbm}
\usepackage{subfigure}
\usepackage{float}
\usepackage{epsfig}
\usepackage{psfrag}
\usepackage{natbib, hyperref}
\usepackage{bm}
\usepackage{amsthm}
\usepackage{eucal}
\usepackage{mathrsfs}
\usepackage{url}
\usepackage{braket}
\usepackage{ulem}
\usepackage{calligra}
\usepackage[T1]{fontenc}
\usepackage[utf8]{inputenc}
\usepackage{newunicodechar}
\usepackage{marvosym}
\usepackage[absolute]{textpos}
\usepackage[cal=cm]{mathalfa}
\usepackage{soul}
\usepackage{lipsum}

\usepackage{color}

\usepackage{hyperref}
\hypersetup{
colorlinks=true,final=true,
        linkcolor=black,
        citecolor=black,
        filecolor=black,
        urlcolor=black,
}
\begin{document}
\setlength{\abovedisplayskip}{3pt}
\setlength{\belowdisplayskip}{3pt}

\newcommand{\pg}{$\varphi_{_{ \rm GS}}$ }
\newcommand{\I}{$I_c$ }

\title{Challenges in detecting topological superconducting transitions via supercurrent \\and phase probes in planar Josephson junctions}
\author{Pankaj Sharma}
\affiliation{Department of Physics, Indian Institute of Technology Roorkee, Roorkee 247667, India}
\author{Narayan Mohanta}
\affiliation{Department of Physics, Indian Institute of Technology Roorkee, Roorkee 247667, India}
\email{narayan.mohanta@ph.iitr.ac.in}


\begin{abstract}
Topological superconductors harbor, at their boundaries and vortex cores, zero-energy Majorana bound states, which can be the building blocks in fault-tolerant topological quantum computing. Planar Josephson junctions host such topological superconducting phases, highly tunable by external magnetic field or phase difference between the superconducting leads. Despite many theoretical and experimental studies, the signatures of the transition to a topological superconducting phase, based on minima in the critical supercurrent $I_c$ flowing across the junction, $0$-$\pi$ transition in the ground state junction phase and their anisotropic magnetic-field response have remained unsettled. Using rigorous numerical calculations with several experimentally-relevant parameter settings, we show that $I_c$ and $\varphi_{_{\rm GS}}$ cannot indicate unambiguously topological transition in any realistic planar junctions. Furthermore, the anisotropic variations of $I_c$ and $\varphi_{_{\rm GS}}$ with in-plane magnetic field appear in junctions that are undoubtedly in trivial superconducting phase, raising concerns on the effectiveness of these probes in identifying topological transitions in  planar junctions. We discuss possible strategies to confirm  a topological superconducting phase in these platforms.
\end{abstract}
      
\maketitle

\section{Introduction}
\begin{figure}[t]
\begin{center}
\vspace{-0mm}
\epsfig{file=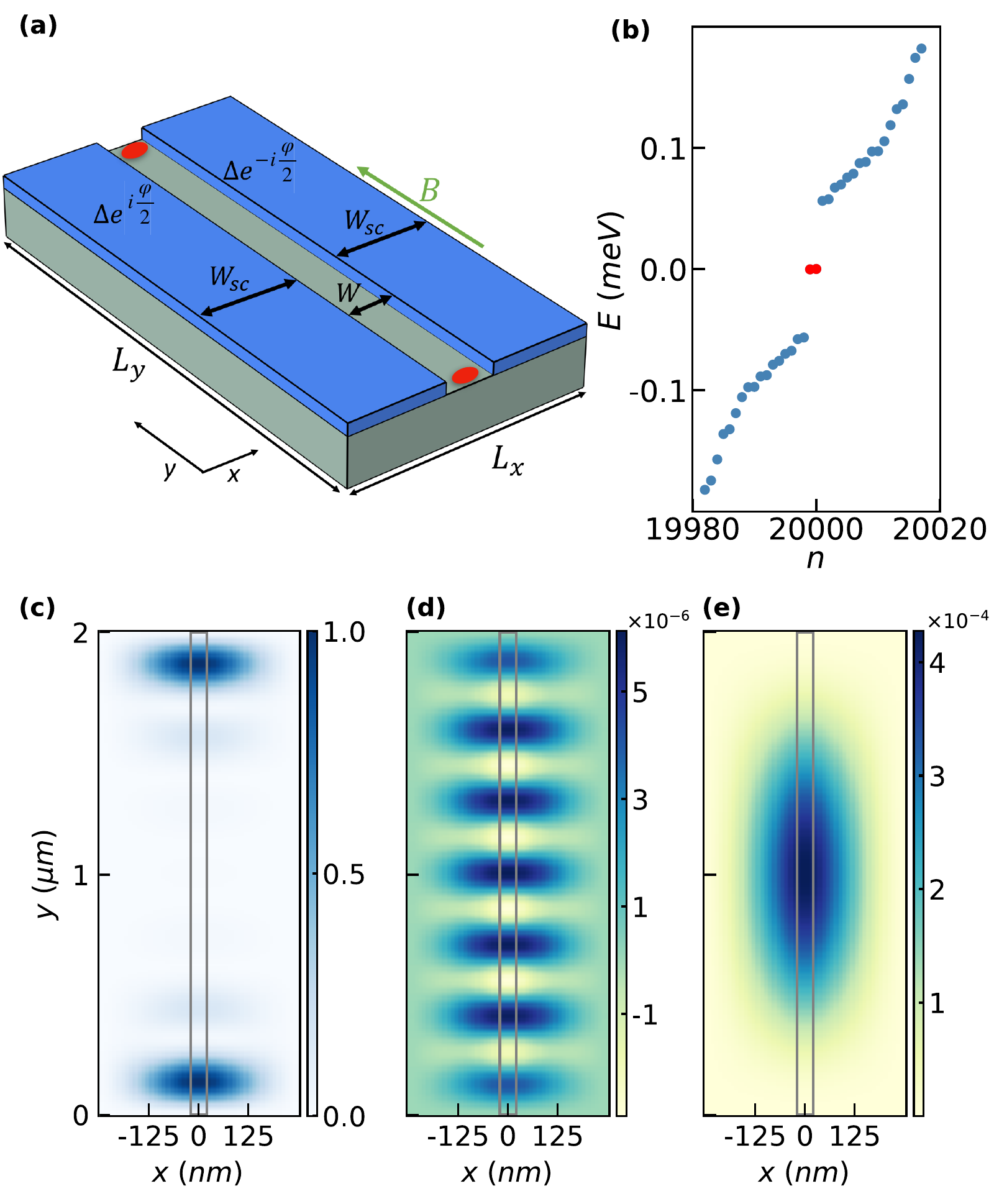,trim=0.0in 0.0in 0.0in 0.0in,clip=false, width=86mm}
\caption{(a) A schematic of the planar Josephson junction, created by placing two superconducting leads on top of a 2DEG based on a semiconducting quantum well. Red markers show the regions in the metallic channel where MBS are localized in the topological phase. (b) Energy eigenvalues versus their index $n$ at $B \!=\! 1.5$~T, $\varphi \!=\! \pi$ and $\mu \!=\!-0.4$~meV, showing two eigenvalues at zero energy (in red), protected from the other states by an energy gap. (c) Local density of state profile (normalized) for the lowest positive energy eigenstate, revealing the localization of the two MBS at the ends of the metallic channel (shown by the grey lines). (d), (e) Profiles of charge density of states at $B \!=\! 1.5$~T (MBS present) and at $B \!=\! 0.5$~T (MBS absent), showing partial charge neutrality in the presence of MBS (colorbar scales show two orders of magnitude reduction in the charge density in (d)). The junction parameters are $L_x\!=\!0.5~\mu$m, $L_y\!=\!2~\mu$m and $W\!=\!0.04~\mu$m.}
\label{fig1}
\vspace{-2mm}
\end{center}
\end{figure}

\vspace{-3mm}
Conclusive identification of a topological superconducting phase hosting zero-energy Majorana bound states (MBS), is currently one of the central problems in condensed matter physics. The MBS, because of their non-Abelian statistics, are promising for building qubits for fault-tolerant topological quantum computing~\cite{Kitaev2000,KITAEV20032,Nayak_2008,Alicea_2011,Alicea_2012}. Extensive efforts in the past years in various geometries with strong Rashba spin-orbit coupling (RSOC), including semiconductor-superconductor nanowires~\cite{Sau_2010,Oreg_2010,Lutchyn_2010,Mourik_Science2012,Das2012,Deng_Science2016,Nichele_PRL2017}, magnetic adatoms-superconductor interfaces~\cite{NadjPerge_PRB2013,NadjPerge_Science2014,Li2016,Feldman2017,Mohanta_PRB2018}, topological insulator-superconductor interfaces~\cite{Fu_PRL2008,Chiu_PRB2011,Xu_PRL2015} and oxide heterostructures~\cite{Mohanta_EPL2014,Kuerten_PRB2017}, revealed zero-bias conductance peak (ZBCP) signature to support the presence of the MBS. However, the challenge has been the disentanglement of the MBS from other low-energy quasiparticles, primarily Andreev bound states and impurity-induced states~\cite{Kells_PRB2012,Prada_PRB2012,Stanescu_PRL2012,Loss_PRB2013,Cayao_PRB2015,DasSarma_PRB2017,Suominen_PRL2017,Vuik_SciPostPhys2019,Pan_PRRes2020,DasSarma_PRB2021}. Therefore, to achieve the sought-after braiding of the MBS, and to overcome the inherent instabilities in one-dimensional nanowire networks, two-dimensional platforms such as the planar Josephson junctions were introduced~\cite{Hell_PRL2017,Pientka_PRX2017,Shabani_PRB2016,Nichele_PRL2020,Wu_PRB2020,Alidoust_PRB2021,Oshima_PRRes2022}.\\
\indent A planar Josephson junction consists of a two-dimensional electron gas (2DEG), proximity-coupled to two superconducting leads (shown schematically in Fig.~\ref{fig1}(a)). It has the additional advantage that the induced topological superconductivity can be easily controlled by the phase difference between the superconducting leads (tunable externally by applying a perpendicular magnetic field). Transition to a topological superconducting phase in these junctions, revealed by the emergence of the zero-energy MBS, appears at a critical value of an in-plane magnetic field, applied along the length of the non-superconducting channel when the phase difference between the superconducting leads is close to $\pi$ ~\cite{Pientka_PRX2017,Hell_PRB2017}. This topological superconducting transition has been suggested to be accompanied in general by a minimum in the critical supercurrent $I_c$ and a jump from nearly $0$ to nearly $\pi$ in the ground state phase $\varphi_{_{\rm GS}}$~\cite{Pientka_PRX2017}. Following this theoretical proposal, successive experiments reported the ZBCP signature of the MBS within a range of magnetic fields, predicted by the minima in $I_c$~\cite{Ren_Nature2019, Fornieri_Nature2019}. Further theoretical work suggested that for narrow superconducting leads ($W_{\rm SC} \lesssim \xi $, where $\xi$ is the superconducting coherence length), as used in the experiments, the minima in $I_c$ do not necessarily indicate topological phase transitions~\cite{Setiawan_PRB2019}. However, a subsequent study, with joint experimental and theoretical efforts, 
concluded a $\pi$ phase jump in the junction phase and a simultaneous minimum in $I_c$ at a critical magnetic field
to be a signature of topological superconducting transition~\cite{Dartiailh_PRL2021}. The conclusions were further supported by the evidence of anisotropic response of $I_c$ and $\varphi_{_{\rm GS}}$ with respect to the in-plane magnetic field. Despite intensive analysis to exclude possible non-topological origins, it remains unclear whether the correspondence between these critical fields can be generalized for different junction dimensions and realistic parameters such as RSOC strength. Moreover, these conflicting results created confusion regarding the viability of these quantities ($I_c$ and $\varphi_{_{\rm GS}}$) as reliable indicators of topological superconducting transition in these planar Josephson junctions.

\indent In this paper, we show that the observables $I_c$, $\varphi_{_{\rm GS}}$, and their anisotropic magnetic-field response cannot predict a topological superconducting transition in any realistic planar Josephson junction, irrespective of $W_{\rm SC} \lesssim \xi$ or $W_{\rm SC} > \xi$. We confirm these findings using the presence (absence) of the zero-energy MBS as the direct indicator of the topological (trivial) phase, in our numerical calculations which were performed using various experimentally-relevant parameter settings and device geometries. Under realistic conditions, $\varphi_{_{\rm GS}}$ exhibits a smooth monotonic variation with in-plane magnetic field rather than a sharp jump. Even though, the critical fields for the minima in $I_c$ may coincidentally match with the critical field for the topological superconducting transition, we could not establish a general correspondence between the two critical fields. We also find that the anisotropy in $I_c$ and $\varphi_{_{\rm GS}}$ with respect to the in-plane magnetic field cannot conclusively distinguish a topological superconducting phase from a trivial one. We discuss possible braiding realization of MBS as a prerequisite to conclude topological superconductivity in a multi-terminal planar junction.

\section{Theoretical Model}
\vspace{-3mm}
The planar Josephson junction devices used in the experiments~\cite{Ren_Nature2019, Fornieri_Nature2019, Dartiailh_PRL2021}, are made of a 2DEG, created at a InAs/HgTe quantum well, and Al superconducting leads. To describe such a 2DEG with proximity-induced superconductivity underneath the superconducting leads, we use the below Hamiltonian
\begin{align} \mathcal{H}&=\sum_{i, \sigma} (4t -\mu)c^\dagger_{i\sigma}c_{i\sigma} - t \sum_{\langle i j\rangle, \sigma} (c^\dagger_{i\sigma}c_{j\sigma} + {\rm H.c.}) \nonumber \\ 
&+\sum_{i } (\Delta_i c^\dagger_{i\uparrow}c^\dagger_{i\downarrow} + {\rm H.c.})   - \dfrac{g^* \mu_{_{\rm B}}B}{2} \sum_{i, \sigma, \sigma^\prime } (\sigma_y)_{\sigma \sigma^\prime}   c^\dagger_{i\sigma} c_{i\sigma^\prime} \nonumber \\ 
&-\dfrac{\mathbf{i}\alpha}{2 a} \sum_{\langle i j\rangle, \sigma \sigma^\prime}    (\bm{\sigma} \times \mathbf{d}_{ij})^z_{\sigma \sigma^\prime}~    c^\dagger_{i\sigma}c_{j\sigma^\prime},
\label{Ham}
\end{align}
where $t\!=\!\hbar^2/2m^*a^2$ is the hopping energy, $m^*$ is the effective mass of the electrons, $a$ is unit spacing of the considered square lattice grid, $i$ and $j$ represent lattice site indices, $\sigma $ and $\sigma^\prime$ are indices for spins ($\uparrow$, $\downarrow$), $\mu$ is the chemical potential, $\Delta_i$ is the superconducting pairing (conventional $s$ wave) amplitude at site $i$, $\mu_{_{\rm B}}$ is the Bohr magneton, $g^*$ is the effective g-factor of the electrons, $B$ is the magnetic field applied along the length of the metallic channel ($y$ direction, in the chosen axes description, as shown in Fig.~\ref{fig1}(a)), $\bm{\sigma}$ represents the Pauli matrices, $\alpha$ is the RSOC strength, $\mathbf{i}$ is the imaginary number, and $\mathbf{d}_{ij}$ denotes the unit vector from site $i$ to $j$. The pairing amplitude is zero in the metallic channel and of constant magnitude on the two superconducting regions, but it has a phase difference of $\varphi$ between the superconducting regions. We use open boundary conditions and the following parameters $m^*\!=\!0.026$ $m_0$ ($m_0$ being the rest mass of electrons), $g^*\!=\!10$, $\Delta \!=\!0.2$~meV, and $\alpha \!=\!30$~meV-nm, typical for Al/InAs systems~\cite{Smith_PRB1987,Mayer_APL2019,Carrad_advmat2020}. We use the lattice grid spacing $a\!=\!10$~nm in our calculations throughout. The eigenvalues and eigenvectors of the Hamiltonian~(\ref{Ham}) were obtained by diagonalizing it using the unitary transformation ${c}_{i \sigma} = \sum_{n}u^n_{i \sigma}{\gamma}_n + v^{n *}_{i \sigma} \gamma^\dagger_n$, where $u^n_{i \sigma}$ ($ v^{n }_{i \sigma}$) represents quasi-particle (quasi-hole) amplitudes, and ${\gamma}_n$ (${\gamma}_n^\dagger$) represents fermionic annihilation (creation) operator of the Bogoliubov-de Gennes quasi-particles corresponding to the $n^{\rm th}$ eigenstate.

\begin{figure*}[ht]
\begin{center}
\vspace{-0mm}
\epsfig{file=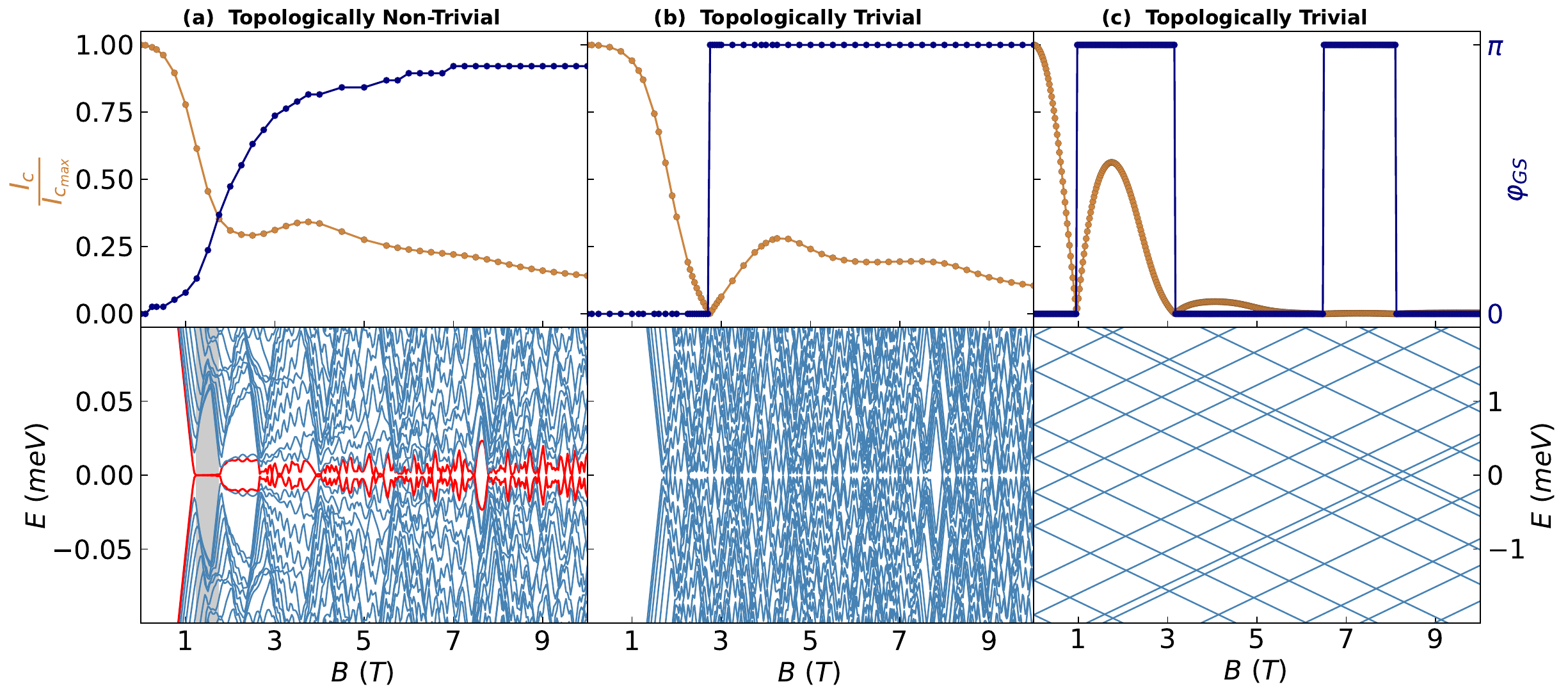,trim=0.0in 0.0in 0.0in 0.0in,clip=false, width= \textwidth}
\caption{Variation with the applied magnetic field $B$ of the critical supercurrent $I_c$ (normalized) and the ground state phase $\varphi_{_{\rm GS}}$ (top row) and quasiparticle energy spectrum (bottom row) for three different cases: (a) a long planar Josephson junction ($L_x\!=\!0.5~\mu$m, $L_y\!=\!2~\mu$m and $W\!=\!0.04~\mu$m) with RSOC strength $\alpha \!=\!30$~meV-nm, (b) the same junction as in (a) but with $\alpha \!=\!0$, (c) a relatively shorter junction ($L_x\!=\!0.3~\mu$m, $L_y\!=\!0.15~\mu$m and $W\!=\!0.1~\mu$m) with $\alpha \!=\!0$. The chemical potential in (a) and (b) is $\mu \!=\!-0.4$~meV and that in (c) is $\mu \!=\!0.5$~meV; the phase difference between the superconducting leads in all three cases is $\varphi \!=\!\pi$. MBS do not appear in the quasiparticle energy spectra in (b) and (c), but $I_c$ exhibits zeros at some critical fields and simultaneously sharp transitions in $\varphi_{_{\rm GS}}$. In case (a), MBS appear (shown by the red lines near zero-energy within the greyed range of $B$), $I_c$ shows a minimum at around $B\!=\!2.5$~T and $\varphi_{_{\rm GS}}$ increases gradually from zero toward $\pi$ with increasing $B$; however, the critical field for the minimum in $I_c$ is different from the critical field at which MBS emerge.}
\label{fig2}
\vspace{-0mm}
\end{center}
\end{figure*}
\begin{figure*}[ht]
\begin{center}
\vspace{-0mm}
\epsfig{file=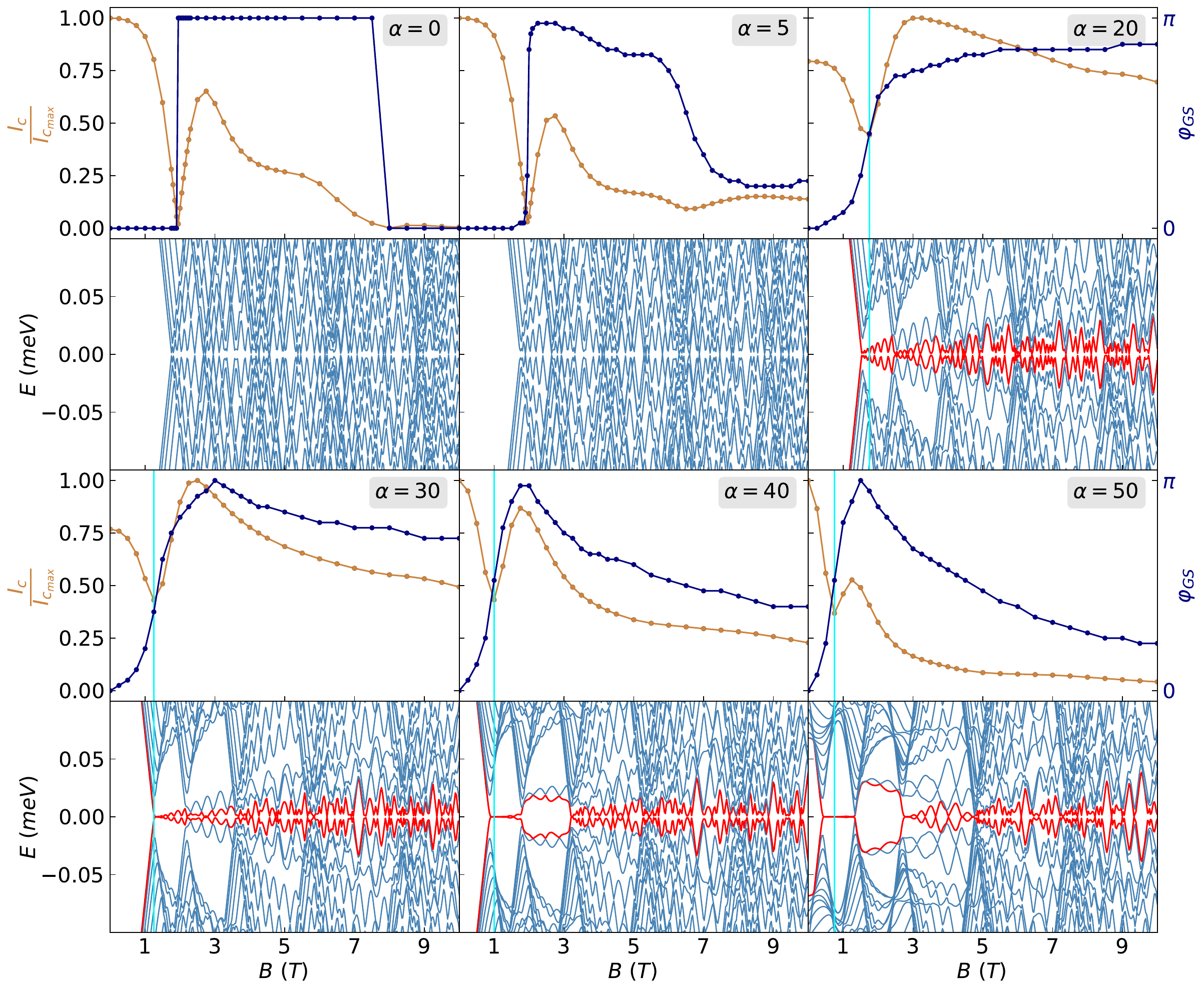,trim=0.0in 0.0in 0.0in 0.0in,clip=false, width=\textwidth}
\caption{Magnetic field-variation of critical supercurrent $I_c$ (normalized) and ground state phase $\varphi_{_{\rm GS}}$ (first and third rows), and corresponding quasiparticle energy spectrum (second and fourth rows) of a planar Josephson junction of dimension $L_x\!=\!0.4~\mu$m, $L_y\!=\!1.6~\mu$m and $W\!=\!0.08~\mu$m, at different RSOC strengths $\alpha$ (in meV-nm unit). Other parameters used are $\varphi \!=\!\pi$ and $\mu \!=\!-0.4$~meV. The cyan vertical lines show quantitative agreement of the critical field for the first minimum in $I_c$ with the critical field for the emergence of zero-energy MBS.}
\label{fig3}
\vspace{-0mm}
\end{center}
\end{figure*}

\section{Results}
\subsection{Appearance of MBS}
\vspace{-3mm}
Zero-energy MBS appear at the two ends of the middle metallic channel of the planar Josephson junction when the phase difference $\varphi$ between the superconducting leads is nearly $\pi$, and the length of the channel $L_y$ is sufficiently large, so that the overlap between the two MBS appearing at its ends is minimized. Other essential requirements are (i) RSOC strength needs to be greater than a critical value, which we found to be nearly $20$~meV-nm, (ii) magnetic field strength $B$ and chemical potential $\mu$ need to be chosen suitably within a range. Quasi-particle energy spectrum for our considered geometry ($L_y\!=\!2~\mu$m, transverse length $L_x\!=\!0.5~\mu$m, channel width $W\!=\!0.04~\mu$m and superconducting lead width $W_{\rm SC} \!=\! 230$~nm (less than the coherence length $\xi \!=\! \hbar v_F/(\pi \Delta)\! \simeq \!885 ~nm$), shown in Fig.~\ref{fig1}(b), reveals presence of a pair of topologically-protected zero-energy MBS at $\varphi \!=\! \pi$, $B \!=\! 1.5$~T, and $\mu \!=\!-0.4$~meV. These zero-energy states are localized near the metallic channel ends, as shown by the local density of states profile in Fig.~\ref{fig1}(c), obtained via $\rho_{i} = \sum_\sigma (|u_{i \sigma}|^2 + |v_{i \sigma}|^2)$ for the lowest positive-energy eigenstate. The local charge density of states $\rho_{ci} = \sum_\sigma (|u_{i \sigma}|^2 - |v_{i \sigma}|^2)$, at two magnetic fields, $B \!=\! 1.5$~T (with MBS) and $B \!=\! 0.5$~T (without MBS) as shown in Figs.~\ref{fig1}(d) and \ref{fig1}(e) respectively, reveals two orders of magnitude reduction in its value, indicating the realization of a partial charge neutrality, supporting further the appearance of the charge-neutral MBS in the junction. Charge density profile in the presence of MBS shows a density-wave-like pattern, a feature that generically appears in all geometries including the nanowire and planar Josephson junction ones, and is a manifestation of oscillatory MBS wave functions~\cite{Mohanta_2021,Mohanta_et_al_2019,supp}. We use these direct confirmations of MBS to identify topological superconducting phase in our geometries.

\subsection{Critical supercurrent and ground state phase}
\vspace{-3mm}
The observables of our main interest are the critical supercurrent $I_c$ and the ground state phase $\varphi_{_{\rm GS}}$, which were predicted to be natural diagnostics for topological transitions in these Josephson junctions~\cite{Pientka_PRX2017}. We calculate the critical supercurrent using $I_c\!=\!{\rm max}\{I(\varphi)\}$, where the supercurrent as a function of the phase difference $\varphi$ is given by the below thermodynamic relation \cite{Beenakker_1992, Beenakker_1992_QPC}
\begin{align}
    I(\varphi) = \dfrac{2e}{\hbar} \dfrac{d{\cal F}}{d\varphi},
\end{align}
and the free energy ${\cal F}$ of the junctions is computed using quasiparticle energies $E_n$ using the relation
\begin{align}
{\cal F} = -2 k_{_B} T \sum_{E_n > 0} \ln \Bigl[ 2 \cosh \Bigl( \dfrac{E_n}{2 k_B T} \Bigr) \Bigr] 
\end{align}
where $k_{_B}$ is the Boltzmann constant and $T$ is the temperature. We use $k_{_B}T\!=\!0.43\Delta$, for the results presented here and explore temperature effect in the Supplemental Material~\cite{supp}. The ground state phase $\varphi_{_{\rm GS}}$ is determined as the phase difference that minimizes ${\cal F}$. We obtain $\varphi_{_{\rm GS}}$ within $0$ and $\pi$, by mapping values appearing above $\pi$ to this range. 

In Fig.~\ref{fig2}(a) (top panel), we show field variations of $I_c$ and $\varphi_{_{\rm GS}}$ for a long junction of dimension and parameters (with finite RSOC) as in Fig.~\ref{fig1}. We find that $I_c$ exhibits a minimum at $B\!\approx\!2.5$~T, accompanied by a gradual increase in $\varphi_{_{\rm GS}}$ from $0$ toward $\pi$ with increasing magnetic field strength $B$. Field-variation of quasiparticle energy spectrum (below panel in Fig.~\ref{fig2}(a)), reveals the emergence of zero-energy MBS within the range $1.1~{\rm T}\!\lesssim \!B\! \lesssim \!1.7~{\rm T}$. Evidently, beyond $B\!\approx \!2.5$~T, the junction is not in topological superconducting phase since MBS are absent. The critical magnetic field $B\!\approx \!2.5$~T, given by the first minimum in $I_c$, is therefore not consistent with the topological phase boundaries. On the other hand, $\varphi_{_{\rm GS}}$ does not reveal any sharp transitions in this case with typical RSOC strengths, available in the discussed 2DEGs, and hence it is also not a good indicator for topological superconducting transitions. Fig.~\ref{fig2}(b) depicts $B$-variation of $I_c$, $\varphi_{_{\rm GS}}$ (top panel) and of quasiparticle energy spectrum (bottom panel) for the same Josephson junction as in Fig.~\ref{fig2}(a) but with zero RSOC. In this case, $I_c$ exhibits a zero and $\varphi_{_{\rm GS}}$ exhibits a sharp jump from $0$ to $\pi$ at a critical field $B\approx2.75$~T; however, the  junction is in trivial phase since MBS do not appear for the entire range of $B$ considered here. We confirm our findings by calculating $I_c$ and $\varphi_{_{\rm GS}}$ also for relatively-shorter junctions; results for a short junction of dimension $L_x\!=\!0.3~\mu$m, $L_y\!=\!0.15~\mu$m and $W\!=\!0.1~\mu$m at zero RSOC are shown in Fig.~\ref{fig2}(c). Multiple ranges of $B$, defined by $\varphi_{_{\rm GS}}\! \approx \! \pi$, can be identified clearly in this case, while the junction remains in trivial superconducting phase. Because of the mean-field nature of the adopted formalism and constant pairing gap $\Delta$, the critical fields are overestimated \textit{i.e.} larger than the actual critical fields found in experiments.

To explore the role of RSOC and variation in the width of the metallic channel in the agreement of the critical fields, we perform similar analyses at different RSOC strengths for a junction with a relatively wider metallic channel (dimension $L_x\!=\!0.4~\mu$m, $L_y\!=\!1.6~\mu$m, $W\!=\!0.08~\mu$m and $W_{\rm SC}\!=\!160$~nm $<\xi$). From the results, shown in Fig.~\ref{fig3}, we note the following: (i) $I_c$ at the first minimum and at the first maximum increases with increasing RSOC strength $\alpha$ till $\alpha \! \approx \!30$~meV-nm, above which it decreases again; such a non-monotonic field dependence exists for shorter junctions also~\cite{supp}, (ii) sharp transitions in $\varphi_{_{\rm GS}}$ appear only for small values of $\alpha$ for which the Josephson junction is in trivial phase always; the gradual increase in $\varphi_{_{\rm GS}}$ with $B$ was also found in Ref.~\cite{Dartiailh_PRL2021}, (iii) the second critical field for the second minimum of $I_c$ or drop in $\varphi_{_{\rm GS}}$ increases with increasing $\alpha$. Therefore, for typical $\alpha$ values, available in the discussed 2DEGs, the second critical field will be beyond the usually-realizable values. The near-zero-energy MBS start to appear above $\alpha \! \approx \!20$~meV-nm with a well-defined energy gap, and the critical field for the first minimum in $I_c$ appears {\it near} the critical field for the topological superconducting transition. Coincidentally, the critical fields are in better agreement for the case of $\alpha \!=\!30$~meV-nm for this junction, than the result presented in Fig.~\ref{fig2}(a) for a junction with a narrower metallic channel. This comparison indicates that the agreement is better for junctions with a wider metallic region. However, there is no general correspondence among these critical fields, particularly for the long junctions considered in the experiments~\cite{Ren_Nature2019,Fornieri_Nature2019,Dartiailh_PRL2021}. With increasing $\alpha$, the critical field for the emergence of MBS decreases, compatible with RSOC-driven topological superconducting transitions. 

\begin{figure}[t]
\begin{center}
\vspace{-0mm}
\epsfig{file=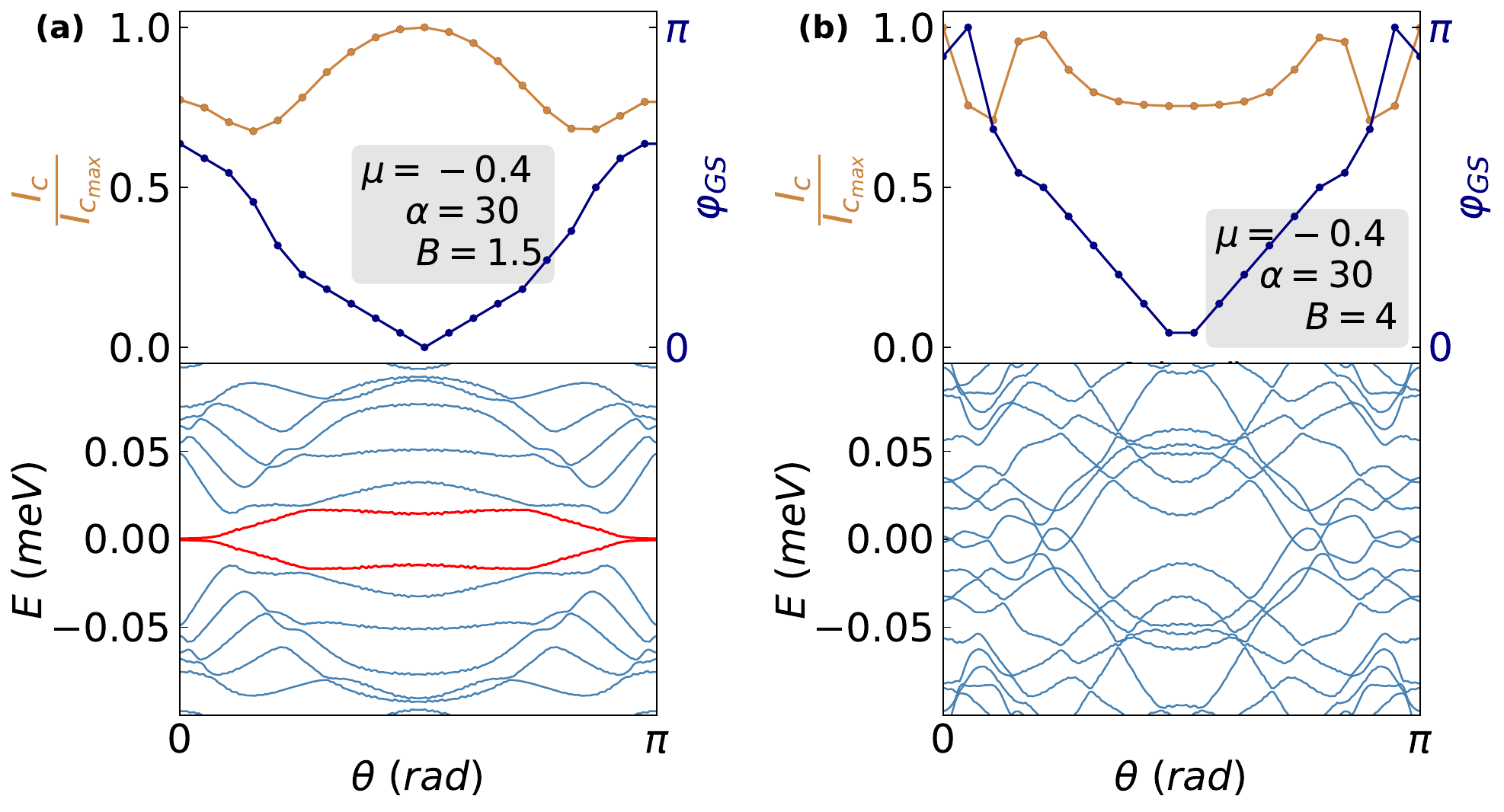,trim=0.0in 0.0in 0.0in 0.0in,clip=false, width=86mm}
\caption{Variation of $I_c$, $\varphi_{_{\rm GS}}$ and the quasi-particle energy spectrum with in-plane magnetic field angle $\theta$ at $\varphi \!=\! \pi$ ($\mu$ in meV unit, $\alpha$ in meV-nm unit and $B$ in T) for two cases: (a) topological phase at $\theta \!=\!0$, and (b) non-topological phase at $\theta \!=\!0$. The system's dimensions are the same as in Fig~\ref{fig3}. Parameters are shown in the insets of the top panels.} 
\label{fig4}
\vspace{-0mm}
\end{center}
\end{figure}

We also investigate the in-plane magnetic field anisotropy in $I_c$, a signature used in Ref.~[\onlinecite{Dartiailh_PRL2021}] in support of topological phase transition. We rotate the in-plane magnetic field angle $\theta$ (measured with respect the $+y$-direction, along the length of the channel) from $\theta \!=\! 0$ to $\theta \!=\! \pi$, and plot $I_c$, $\varphi_{_{\rm GS}}$ and the energy spectrum in Fig.~\ref{fig4}. We show two cases: (a) $B\!=\!1.5$~T (topological phase at $\theta \!=\! 0$ as shown in Fig.~\ref{fig3}) (b) $B\!=\!4$~T (non-topological phase at $\theta \!=\!0$ as shown in Fig.~\ref{fig3}), at a fixed RSOC strength $\alpha \!=\! 30$~meV-nm. It is evident that the magnetic anisotropy in $I_c$ and $\varphi_{_{\rm GS}}$ exists in both cases. There is no noticeable difference in the field variation of $I_c$ and $\varphi_{_{\rm GS}}$ (see also Fig.S3 in Ref.~\onlinecite{supp}). Hence, further evidence such as the demonstration of the non-Abelian characters of the MBS is required, in addition to the signatures in $I_c$ and $\varphi_{_{\rm GS}}$, in order to unambiguously detect a topological superconducting transition in these planar junctions.\\

\section{Discussion and outlook}
\vspace{-3mm} 
To generalize our findings, we performed similar calculations of $I_c$ and $\varphi_{_{\rm GS}}$, as presented above for various other scenarios: (i) planar junctions with wider superconducting leads \textit{i.e.} $W_{\rm SC} \!>\! \xi$ and wider channels, (ii) magnetic field applied only in the metallic channel region, (iii) non-uniform $g$ factor~\cite{supp}. Based on these extensive analyses of different parameters and device geometries, we conclude that, with realizable device dimensions and realistic parameters, $I_c$ and $\varphi_{_{\rm GS}}$ cannot identify topological superconducting transitions in planar Josephson junctions.

Other observables used in previous experiments include (i) fractional ac Josephson effect and missing Shapiro steps due to $4\pi$ periodic current-phase relation~\cite{Kwon2004,Aguado_PRL2012,Sau_PRB2017,Rokhinson2012,Wiedenmann2016,Deacon_PRX2017}, (ii) ZBCP signature~\cite{Ren_Nature2019,Fornieri_Nature2019,Banerjee_PRL2023,Banerjee_PRB2023}, (iii) bulk gap closing signature ~\cite{Banerjee_PRL2023,Banerjee_PRB2023}, (iv) local and non-local transport spectroscopy~\cite{Banerjee_PRL2023}, and (v) negative conductance curvature at zero bias $(\partial^2 G/\partial V^2)|_{V=0}$~\cite{Fornieri_Nature2019}. These signatures likely cannot confirm topological superconductivity separately due to the presence of non-Majorana states~\cite{Dartiailh2021,Rosenbach_SciAdv2021}, but together may constitute a procedure for the detection of the topological superconductivity.

To look forward, it is intuitively appealing to explore more complicated geometries such as multi-terminal planar Josephson junctions~\cite{Riwar2016, Pankratova_PRX2020} for realizing Majorana fusion and non-Abelian Majorana braiding~\cite{Flensberg_PRL2011,van_Heck_2012,Hyart_PRB2013,Hell_PRB2017,Stern_PRL2019,Bradraj_PRB2023}. These complicated geometries may come with new challenges, one of them clearly being the issue of fixing the direction of the in-plane magnetic field which is required to be along the metallic channel length. This particular problem can be overcome by placing underneath the Josephson junction a chiral magnetic texture such as a skyrmion crystal which can provide a gauge field and create a two-dimensional topological superconductivity in the entire 2DEG~\cite{Desjardins2019,Mohanta_2021}. Nonetheless, these planar Josephson junctions provide a versatile two-dimensional platform, capable of manipulating MBS with more efficient control knobs~\cite{Laeven_PRL2020,Alidoust_PRB2018,Scharf_PRB2019,Moehle_NanoLett2021,Paudel_PRB2021}, and it is possible to realize even more exotic states~\cite{Klinovaja_PRB2014,vonOppen_PRB2017}.

\section*{Acknowledgements}
\vspace{-3mm} 
PS was supported by Ministry of Education, Government of India via a research fellowship. NM acknowledges support of initiation grant (No. IlTR/SRIC/2116/FIG) from IIT Roorkee. Numerical calculations were performed at the computing resources of PARAM Ganga at IIT Roorkee, provided by National Supercomputing Mission, implemented by C-DAC, and supported by the Ministry of Electronics and Information Technology and Department of Science and Technology, Government of India.

\vspace{-0mm}
\vspace{-2mm}

%


\newpage
\pagebreak
\clearpage
\widetext
\begin{center}
\textbf{{Supplementary Information: "Challenges in detecting topological superconducting transitions via supercurrent and phase probes in planar Josephson junctions"}}
\vspace{1em}\\
{Pankaj Sharma and Narayan Mohanta}\\
\textit{\small Department of Physics, Indian Institute of Technology Roorkee, Roorkee 247667, India}

\end{center}
\setcounter{equation}{0}
\setcounter{figure}{0}
\setcounter{table}{0}
\setcounter{page}{1}
\makeatletter
\renewcommand{\theequation}{E\arabic{equation}}
\renewcommand{\thefigure}{S\arabic{figure}}
\renewcommand{\bibnumfmt}[1]{[R#1]}
\renewcommand{\citenumfont}[1]{R#1}

\begin{figure*}[ht]
\begin{center}
\vspace{-0mm}
\epsfig{file=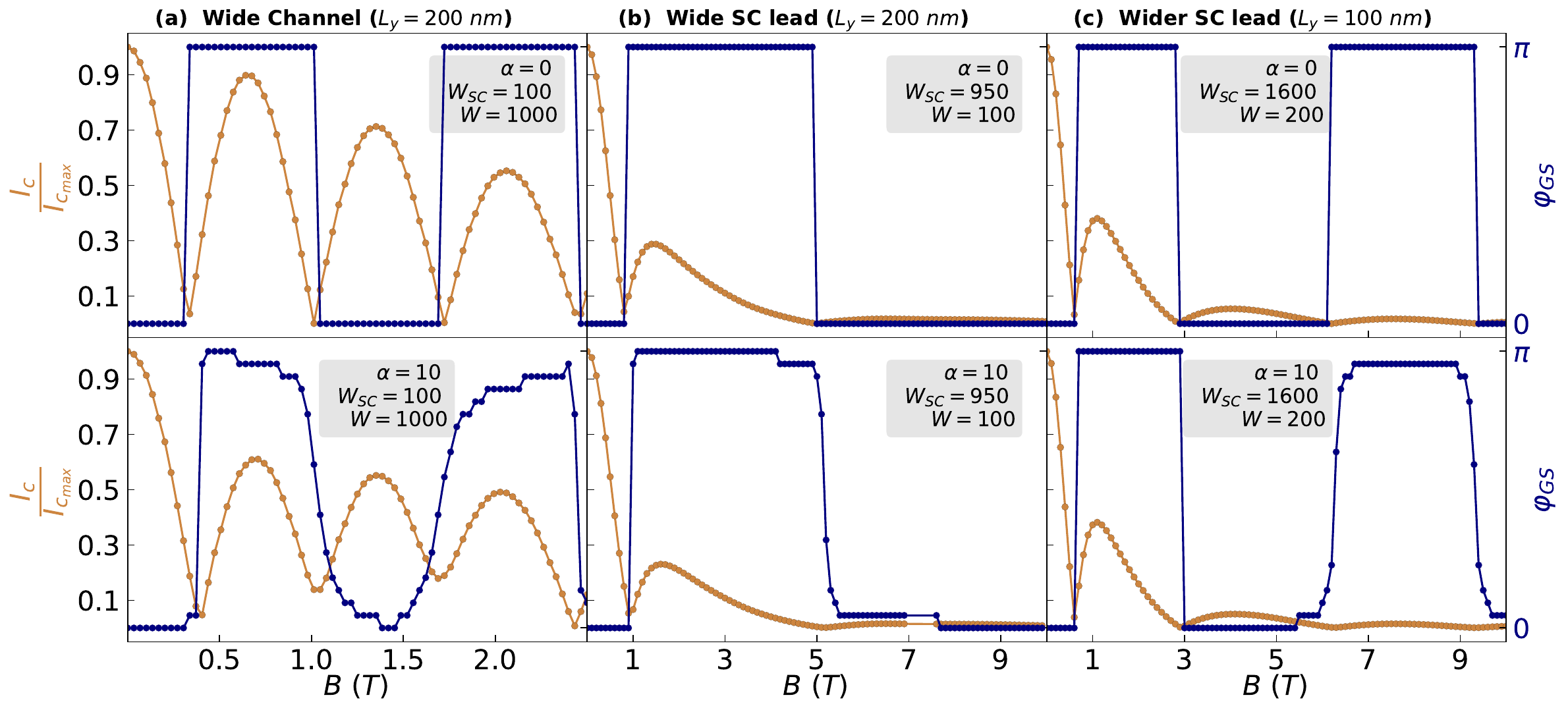,trim=0.0in 0.0in 0.0in 0.0in,clip=false, width=\textwidth}
\caption{Magnetic field variation of critical supercurrent (\(I_c\)) and ground state phase (\(\varphi_{_{\rm GS}}\)) of planar Josephson junctions of different dimensions and different RSOC strengths ($\alpha$ in units of meV-nm) and fixed chemical potential ($\mu = 5$~meV-nm): (a) wide channel and narrow superconducting leads ($W > \xi$, $W_{\rm SC} \ll \xi$, $W$ and $W_{\rm SC}$ in units of $nm$), (b) narrow channel and wide superconducting leads ($W \ll \xi$, $W_{\rm SC} \gtrsim \xi$), (c) narrow channel and wider superconducting leads ($W \ll \xi$, $W_{\rm SC} > \xi$) with $W_{\rm SC}$ approximately twice of the superconducting coherence length $\xi$. }
\label{fig:S1}
\vspace{-4mm}
\end{center}
\end{figure*}

\begin{figure*}[ht]
\begin{center}
\vspace{-0mm}
\epsfig{file=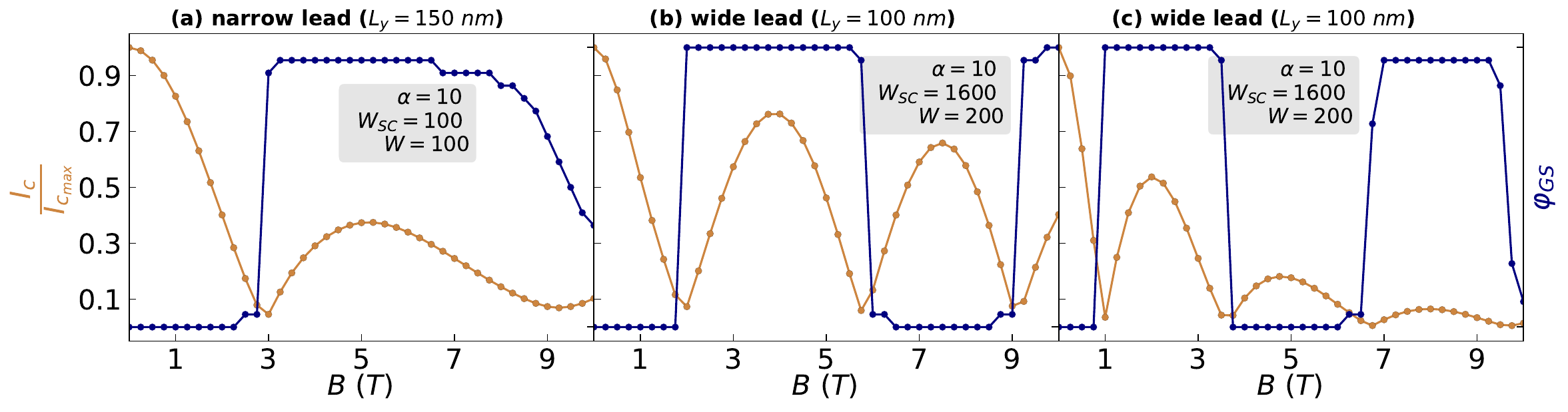,trim=0.0in 0.0in 0.0in 0.0in,clip=false, width=\textwidth}
\caption{Magnetic field variation of critical supercurrent (\(I_c\)) and ground state phase (\(\varphi_{_{\rm GS}}\)) of planar Josephson junctions of different dimensions: (a),(b) when the field is applied only in the metallic channel, or (c) when the $g$ factor is different in different regions.  In panel (a), the junction dimensions and other parameters that are not specified are the same as in Fig.2(c) of the main text ($W$ and $W_{\rm SC}$ in nm unit). In panel (b), the dimensions and parameters are the same as in Fig.~\ref{fig:S1}(c). In panel (c), the parameters used are $g \!=\! 20$ in the superconducting leads and $g \!=\! 50$ in the metallic channel.}
\label{fig:S2}
\vspace{-4mm}
\end{center}
\end{figure*}

\begin{figure*}[ht]
\begin{center}
\vspace{-0mm}
\epsfig{file=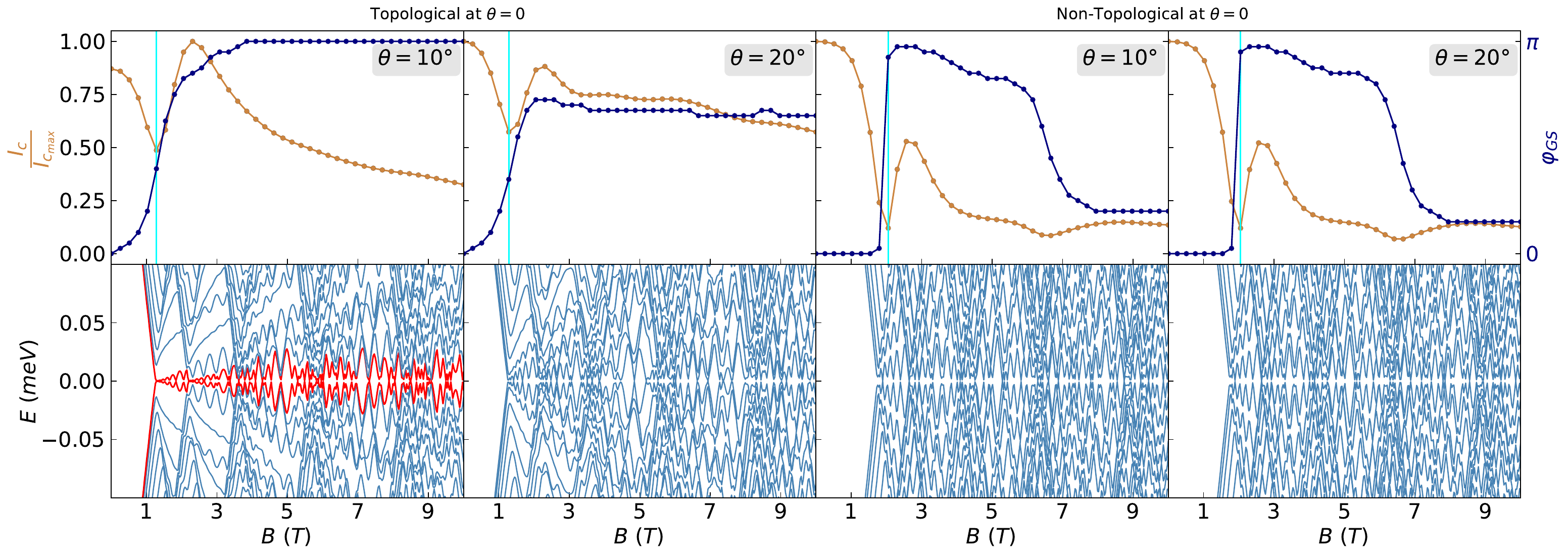,trim=0.0in 0.0in 0.0in 0.0in,clip=false, width=\textwidth}
\caption{Magnetic field variation of critical supercurrent (\(I_c\)) and ground state phase (\(\varphi_{_{\rm GS}}\)) at different field angles $\theta$. The first two columns are for $\theta \!=\! 10^{\circ}$ and $\theta \!=\! 20^{\circ}$ at a fixed RSOC strength $\alpha \!=\!30$~meV-nm which supports MBS at $\theta \!=\! 0 ^{\circ}$ (as shown in Fig.~3 of the main text). The third and fourth columns are for $\theta \!=\! 10^{\circ}$ and $\theta \!=\! 20^{\circ}$ at a fixed RSOC strength $\alpha \!=\!5$~meV-nm which does not support MBS at $\theta \!=\! 0 ^{\circ}$ (as shown in Fig.~3 of the main text). All other parameters are the same as in Fig.~3 of the main text.}
\label{fig:S3}
\vspace{-4mm}
\end{center}
\end{figure*}

\begin{figure}[ht]
\begin{center}
\vspace{-0mm}
\epsfig{file=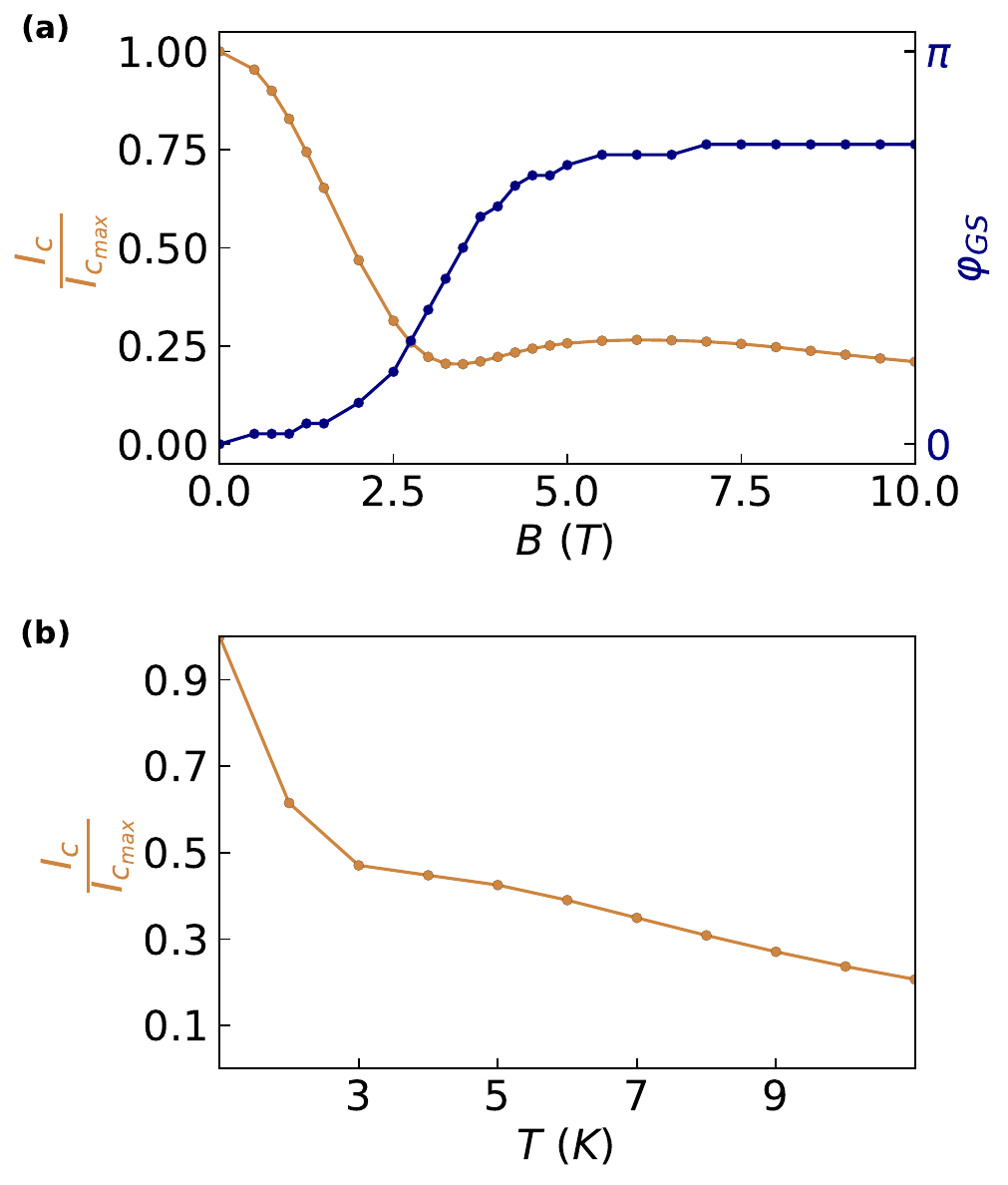,trim=0.0in 0.0in 0.0in 0.0in,clip=false, width=86mm}
\caption{ (a) Critical supercurrent (\(I_c\)) and ground state phase (\(\varphi_{_{\rm GS}}\)) with varying in-plane magnetic field (\(B\)) in a long planar Josephson junction with a narrow channel (same as the one used in Fig 2(a) in the main paper) at a temperature \(k_B T\! \simeq \!1.29\Delta\); \(I_c\) shows a minimum around $B\!=\!3.5$~T, while the ground state phase (\(\varphi_{_{\rm GS}}\)) gradually increases from nearly zero to nearly 0.75\(\pi\). (b) Temperature variation of \(I_c\) at a fixed magnetic field $B\!=\!2.75$~T, showing a monotonic decrease in \(I_c\) with increasing temperature.}
\label{fig:S5}
\vspace{-4mm}
\end{center}
\end{figure}

\noindent {\bf I. Different junction dimensions:}\\
In Fig~\ref{fig:S1}, we present the variations of the critical supercurrent (\(I_c\)) and the ground state phase (\(\varphi_{_{\rm GS}}\)) for different dimensions of  planar Josephson junctions. We consider both narrow superconducting leads \textit{i.e.} $W_{\rm SC} \! \lesssim \! \xi$, where $\xi$ is the superconducting coherence length (found out to be about 885 nm for the considered junctions), and wide superconducting leads \textit{i.e.} $W_{\rm SC} \! > \! \xi$. For these considered junctions, a topological phase does not appear, as the junctions do not support the Majorana bound states (MBS). The description in Ref.~[\onlinecite{Setiawan_PRB2019}] for the limit $W_{\rm SC} \! \lesssim \! \xi$ suggested that $I_c$ does not unambiguously reveal a topological superconducting transition. Here, we find that for both these limits, \textit{i.e.} for junctions of any dimensions, both $I_c$ and $\varphi_{_{\rm GS}}$ do not indicate a topological phase transition.

\noindent {\bf \\II. Non-uniform magnetic field and $g$ factor:}\\
In Fig~\ref{fig:S2}, we show the field variation of \I and \pg when the external field is applied only in the metallic channel and also when the planar Josephson junction has different $g$-factors in the metallic channel and superconducting leads.
We find that the first minimum in \I occurs at a larger value of the magnetic field when the field is applied only in the middle channel; this is true for both narrow and wide leads as shown in Fig.~\ref{fig:S2}(a) and Fig.~\ref{fig:S2}(b), respectively, without the trace of any topological transition. No distinct behavior is found for the case when the $g$ factor in the superconducting leads is less than that in the metallic region, as shown in Fig.~\ref{fig:S2}(c).


\noindent {\bf \\III. Rotation of in-plane magnetic field:}\\
In Fig~\ref{fig:S3}, the in-plane field variation at different field angles $\theta$ (measured with respect to the $+y$ direction) is shown for two cases: (i) $\alpha \!=\!30$~meV-nm (topological phase at $\theta \!=\!0$ as in Fig.~[3] of the main text), (ii) $\alpha \!=\!5$~meV-nm (non-topological phase at $\theta \!=\!0$ as in Fig.~[3] of the main text). We find that the case (i) at $\theta \!=\!10^{\circ}$ still supports zero-energy MBS. At $\theta \!=\! 20^{\circ}$ the MBS disappear. But both panels (first and second columns of Fig.~\ref{fig:S3}) show minima in \I at a critical magnetic field, and corresponding gradual increase in \pg from nearly $0$ to nearly $\pi$ for both $\theta \!=\!10^{\circ}$ and $\theta \!=\!20^{\circ}$. The non-topological case (ii) (third and fourth columns of Fig.~\ref{fig:S3}) also show minima in \I and $0-\pi$ jumps in \pg without showing any zero energy MBS at different tilt angles of the applied field. Even if the applied field is tilted, the minima in \I occurs at around the same magnitude of the in-plane field for topological and non-topological cases at $\theta \!=\!0$. The above analysis suggests that the tilting of the in-plane field away from the channel-length direction destroys any existing topological superconducting phase, but the minima in \I and $0-\pi$ jumps in \pg prevail for topological as well as non-topological cases.\\

\noindent {\bf IV. Temperature variation:}\\
In Fig \ref{fig:S5}(a), the behavior of \I and \pg is presented for a long junction ($L_y$ large) with a narrow channel ($W_{\rm SC}$ small), at a temperature, given by $k_{_B} T\! \simeq \!1.29\Delta$, approximately three times the value used in the results in the main paper. The critical supercurrent exhibits its first minimum around a magnetic field value of $3.5$~T, whereas, in Fig.~2(a) of the main paper, this minimum was observed around $2.5$~T. This shift to a higher magnetic field strength for the first minimum in \I is because of a higher temperature of the junction.
On the other hand, \pg gradually increases from nearly zero and saturates at a value close to $\pi$ at this elevated temperature also. However, the deviation of this saturation value of \pg from $\pi$ increases with increasing temperature. These calculations indicate that the first minimum in \I occurs at a higher magnetic field strength, and its value at the higher critical field decreases as the temperature increases.
In Fig.~\ref{fig:S5}(b), we show temperature variation of \(I_c\) for the considered long junction with a narrow channel. As temperature increases, \(I_c\) decreases monotonically, revealing a shoulder-like feature at around 3~K. Temperature variation of  \(I_c\), observed experimentally in a planar S-N-S geometry~\cite{golikova_critical_2013}, also shows a monotonic decrease. In our case, \(I_c\) does not approach zero even for temperatures exceeding \(10\)~K, contrary to what is reported in this experimental observations. This deviation is because of a constant pairing gap used in our simulations. Interestingly, a similar shoulder-like feature was also seen in the experiments~\cite{golikova_critical_2013}.

\end{document}